# Vulnerabilities in Quantum Key Distribution Protocols

D. Richard Kuhn
kuhn@nist.gov
National Institute of Standards and Technology

**Abstract.** Recently proposed quantum key distribution protocols are shown to be vulnerable to a classic man-in-the-middle attack using entangled pairs created by Eve. It appears that the attack could be applied to any protocol that relies on manipulation and return of entangled qubits to create a shared key. The protocols that are cryptanalyzed in this paper were proven secure with respect to some eavesdropping approaches, and results reported here do not invalidate these proofs. Rather, they suggest that quantum cryptographic protocols, like conventional protocols, may be vulnerable to methods of attack that were not envisaged by their designers.

**Introduction.** The history of cryptography is replete with examples of protocols that were believed to be secure but shown to be vulnerable to novel attacks, often years after their design. Although sophisticated tools (and even specialized logics) have been designed to analyze and prove various properties of protocols, it is generally accepted that the best assurance of security is obtained through careful review by many experts familiar with vulnerabilities of similar protocols. An experience base of known vulnerabilities is a crucial component of this review. Because it is so new that there were no commercial products available before this year, little is known of vulnerabilities that may occur in the design and implementations of quantum cryptographic protocols. But if the history of cryptography is a reliable guide, it should be expected that even quantum cryptographic protocols designed by experts will have unanticipated vulnerabilities. The results presented in this paper suggest that this expectation is as true for quantum cryptography as it is for conventional.

**Protocol Vulnerabilities.** Li [1] describes a QKD protocol using Greenberger-Horne-Zeilinger (GHZ) states that requires no classical communication. The protocol is described as follows, for communicating parties Alice and Bob:

1. Alice creates a three qubit system in GHZ state $\frac{1}{\sqrt{2}}(|000\rangle + |111\rangle)$, sending the third qubit to Bob.
2. To encode a '1', Bob uses the operator $s_x$ on the received qubit; to encode '0', he does nothing to the received qubit.
3. Bob returns the qubit to Alice.
4. Steps 1 – 3 are repeated, with Alice combining each received qubit with the corresponding two qubits of the original tripartite systems she retained, until a bit stream has been received by Alice. Alice then executes a controlled-NOT operation on the first two qubits, with the second qubit as control and the first as target. She then does a Bell state measurement on the last two qubits (of the 3-qubit GHZ system). She then maps the Bell state measurements as follows:

$$\frac{1}{\sqrt{2}}(|00\rangle + |11\rangle) = \text{'0'}; \quad \frac{1}{\sqrt{2}}(|01\rangle + |10\rangle) = \text{'1'}$$

A Bell state of $\frac{1}{\sqrt{2}}(|00\rangle - |11\rangle)$ or $\frac{1}{\sqrt{2}}(|01\rangle - |10\rangle)$ indicates eavesdropping by Eve.

Li [1] shows that this protocol is secure with respect to an attack in which Eve measures qubits returning from Bob to Alice, with a probability that Eve escapes detection of $2^{-n}$, for $n$ qubits. It is also shown that the protocol is secure with respect to an attack where Eve executes a controlled-NOT operation on the qubits sent from Bob to Alice.

Unfortunately, the protocol is vulnerable to a quantum version of a classic man-in-the-middle attack, which we will refer to as an *EPR man-in-the-middle attack*, conducted as follows:

1. Alice creates a three-qubit system in GHZ state $\frac{1}{\sqrt{2}}(|000\rangle + |111\rangle)$, sending the third qubit to Bob.
2. Eve captures the qubit, creates her own two-qubit system, then forwards to Bob one qubit of a two-qubit system in EPR state $\frac{1}{\sqrt{2}}(|00\rangle + |11\rangle)$.
3. To encode a '1', Bob uses the operator $\sigma_x$ on the received qubit; to encode '0', he does nothing to the received qubit.
4. Bob returns the qubit to Eve, thinking it is being returned to Alice.
5. Eve combines the received qubit with the one she retained from the EPR pair that she created, then executes a Bell state measurement on the pair. Bit values are decoded as in step 4 of the Li protocol. Eve then records the bit value for the qubit received from Bob Taking the qubit she captured previously from Alice, she either executes $s_x$ to encode a '1' or does nothing to encode '0', and returns the qubit to Alice. At end, Eve has a complete copy of the key shared by Alice and Bob.

The attack requires Eve to know or guess the basis that is used by Alice and Bob, but since no classical communication is exchanged, the basis must be the same throughout the protocol. In a realistic implementation, the basis will be either standard, or chosen from a small number of possibilities that Eve can guess and determine quickly in a high traffic network. Since any realistic implementation will have less than perfect transmission, Eve can evade detection by removing qubits at a low enough rate to remain below the normal transmission error rate. Eve can measure random qubits in a basis of her choosing, obtaining a long-term distribution of values for 0 and 1 in her basis, designated $P_0 = \alpha^2$ and $P_1 = \beta^2$ for the proportion of 0 and 1 respectively. If she has guessed correctly, she will obtain $P_0 = P_1 = .50$. If not, she simply rotates her basis enough to obtain the 50/50 distribution, an angle of $q = (\cos^{-1}\sqrt{P_0}) - p/4$. This relationship can be seen from Figure 1.

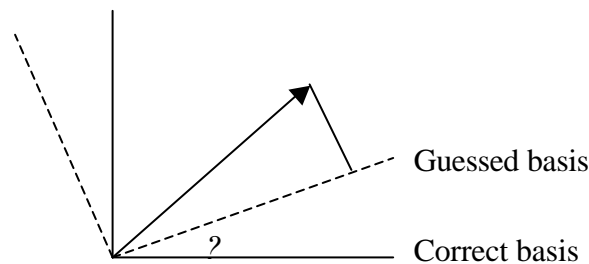

Figure 1. Determining correct basis from long-term distribution.

Boström and Felbinger [2] describe a secure communication protocol based on entangled pairs of qubits. The protocol is described as follows, for communicating parties Alice and Bob:

> p.0) Protocol is initialized. $n = 0$. The message to be transmitted is a sequence $x^N = (x_1,..., x_N)$, where $x_n \in \{0,1\}$.
>
> p.1) $n = n + 1$. Alice and Bob are in message mode. Alice prepares two qubits in the Bell state $\frac{1}{\sqrt{2}}(|01\rangle + |10\rangle)$
>
> p.2) She stores one qubit, the *home qubit*, and sends the other one, the *travel qubit*, to Bob through the quantum channel.
>
> p.3) Bob receives the *travel qubit*. With probability $c$ he switches to control mode and proceeds with $c.1$, else he proceeds with $m.1$.
>
>> c.1) Bob measures the travel qubit in the basis $B_z = \{0, 1\}$ and obtains the result $i \in \{0,1\}$ with equal probability.
>>
>> c.2) He sends $i$ through the public channel to Alice.
>>
>> c.3) Alice receives $i$ from the public channel, switches to control mode and measures the home qubit in the basis $B_z$ resulting in the value $j$.
>>
>> c.4) ($i = j$): Eve is detected. Abort transmission.
>> ($i \ne j$): Set $n = n - 1$ and go to p.1.
>
>> m.1) To encode a '1', Bob uses the operator $s_z$ on the received qubit; to encode '0', he does nothing to the received qubit, then returns the qubit to Alice.
>>
>> m.2) Alice receives the travel qubit, combines it with the qubit she retained, and does a Bell measurement on the pair. The measurement is then decoded as:
>>
>> $$\frac{1}{\sqrt{2}}(|01\rangle + |10\rangle) = 0; \quad \frac{1}{\sqrt{2}}(|01\rangle - |10\rangle) = 1.$$
>>
>> m.3) ($n < N$): Go to p.1. ($n = N$): Go to p.4.
>
> p.4. Message $x^N$ is transmitted from Bob to Alice. Communication successfully terminated.

This protocol is also vulnerable to the EPR man-in-the-middle attack under the following conditions:
- Eve can read and modify the quantum channel, including the ability to capture qubits.
- Eve can read and modify the classical channel. In the protocol, this channel is referred to as a *public* classical channel, but for a realistic implementation in a network, it is very unlikely that it will be public, so without additional protocol steps as safeguards, Eve's modification would not be observable by either Alice or Bob.

The attack proceeds as follows:
- At step p.2, Eve captures the qubit, then forwards to Bob one qubit of a two-qubit system that she has created in EPR state $\frac{1}{\sqrt{2}}(|00\rangle + |11\rangle)$.
- At step m.1, Bob encodes '0' or '1' using $s_z$ then returns the qubit to Eve, thinking it is being sent to Alice. Eve does a Bell state measurement to decode the received qubit: state $\frac{1}{\sqrt{2}}(|00\rangle + |11\rangle)$ is decoded as '0'; state $\frac{1}{\sqrt{2}}(|00\rangle - |11\rangle)$ is decoded as '1'. She executes either $s_z$ (for '1') or does nothing (for '0') to the captured qubit to encode the bit sent by Bob, then forwards it to Alice.
- At step c.1, Bob measures the travel qubit, then sends this result on the classical channel. Eve intercepts the transmission, then measures her captured qubit to obtain either a '0' or '1' with equal probability. She then forwards this value to Alice.

Eve must have the ability to intercept the control (eavesdropping detection) bits sent from Bob to Alice, measure the received qubit and substitute a value to be returned to Alice in real time. Compared with attacks on classical cryptographic protocols, these assumptions are fairly minimal, and not at all unrealistic. The examples presented in this section illustrate that quantum cryptographic protocols may be vulnerable to many of the same attacks as classical protocols. Furthermore, as with conventional protocols, quantum protocols may be vulnerable due to subtle design issues. A protocol that is similar to ping-pong coding, developed by Beige, Englert, Kurtsiefer, and Weinfurter [3], is not vulnerable to the EPR man-in-the-middle attack because the location of the control bits is not known until completion of qubit transmission. Therefore Eve could not determine which bits to measure and what value to substitute and return to Alice.

A protocol developed by Cai [4] uses the ping-pong protocol without a public channel. The Cai protocol attempts to add security by randomizing the Bell state of the receiving party, rather than using a consistent state as in Bostrom and Feilberger. To transmit a bit, Alice selects at random from one of two Bell states: $\Psi^+ = \frac{1}{\sqrt{2}}(|01\rangle + |10\rangle)$ or $\Phi^+ = \frac{1}{\sqrt{2}}(|00\rangle + |11\rangle)$. The protocol proceeds as follows:

1. Alice prepares two qubits in either of the two states $\Psi^+$ or $\Phi^+$, selected randomly.
2. Alice sends one of the qubits from the EPR pair to Bob and retains the other.
3. To encode a '1', Bob uses the operator $s_z$ on the received qubit; to encode '0', he does nothing to the received qubit.
4. Bob returns the qubit to Alice.
5. Alice receives the travel qubit, combines it with the qubit she retained, and does a Bell measurement on the pair. If the state of the EPR pair created in Step 1 was $\Psi^+$ then the measurement is then decoded as:

$$\frac{1}{\sqrt{2}}(|01\rangle + |10\rangle) = 0; \quad \frac{1}{\sqrt{2}}(|01\rangle - |10\rangle) = 1.$$ If the result of the Bell measurement is $\Phi^+$ or $\Phi^-$, then Eve is detected.

If the EPR pair state was $\Phi^+$, then the decoding step is

$$\frac{1}{\sqrt{2}}(|00\rangle + |11\rangle) = 0; \quad \frac{1}{\sqrt{2}}(|00\rangle - |11\rangle) = 1.$$ If the result of the Bell measurement is $\Psi^+$ or $\Psi^-$, then Eve is detected.

Like the protocols described previously, Eve can obtain a perfect copy of the message transmitted from Bob to Alice. The attack is essentially the same:
- At step 2, Eve captures the qubit, then forwards to Bob one qubit of a two-qubit system that she has created in EPR state $\frac{1}{\sqrt{2}}(|00\rangle + |11\rangle)$.
- At step 3, Bob encodes '0' or '1' using $s_z$ then returns the qubit to Eve, thinking it is being sent to Alice. Eve does a Bell state measurement to decode the received qubit: state $\frac{1}{\sqrt{2}}(|00\rangle + |11\rangle)$ is decoded as '0'; state $\frac{1}{\sqrt{2}}(|00\rangle - |11\rangle)$ is decoded as '1'. She executes either $s_z$ (for '1') or does nothing (for '0') to the captured qubit to encode the bit sent by Bob, then forwards it to Alice.

Although the Cai protocol appears at first to be secure because the Bell state is chosen randomly in Step 1, this attack needs fewer assumptions to succeed than the attack on the ping-pong protocol. Eve needs only the ability to intercept qubits and replace them with hers, and to determine the basis used by Alice, which can be done using a relatively small number of intercepted qubits using the procedure described earlier.

**Discussion and Conclusions.** The EPR man-in-the-middle attack works because Alice and Bob cannot detect that qubits received by Bob are part of an entangled pair sent by Eve. Although the two protocols use different sigma operators (*X* and *Z* respectively), Eve can duplicate the actions of Bob without detection, regardless of the operator used by Bob. In the first case, this occurred because of the lack of conventional communication, and in the second because Eve was assumed able to control the classical communication between Alice and Bob, and the protocol did not provide steps for them to authenticate messages. Adapting methods described by Vaidman et al. [5], it would always be possible for Eve to distinguish the sigma operator applied by Bob. It thus appears that any protocol similar to the ones described above would be vulnerable.

**Acknowledgments.** I am grateful to Carl Williams for suggesting significant improvements to the original draft of this paper.

**References**
[1] Li, X., "A quantum key distribution protocol without classical communication," quant-ph/0209050 (September 6, 2002).

[2] K. Boström and T. Felbinger, "Ping-pong Coding", quant-ph 0209040 (September 5, 2002).

[3] A. Beige, B.-G. Englert, C. Kurtsiefer, and H. Weinfurter, "Secure Communication with a Publicly Known Key", Phys. Pol. A 101, 357 quant-ph 0111106, May 9, 2002.

[4] Cai, Q-Y. "Deterministic Secure Direct Communication Using Ping-pong Protocol without Public Channel", quant-ph 0301048, January 13, 2003.

[5] How to Ascertain the Values of Sigma_x, Sigma_y, and Sigma_z of a Spin-1/2 Particle L. Vaidman, Y. Aharonov and D. Albert. *Phys. Rev. Lett. 58, 1385 (1987)*